\documentclass[aps,amsfonts,amsmath,prl,twocolumn]{revtex4}
\usepackage{epsf,mathrsfs,bbm}
\usepackage{graphicx} 
\usepackage{xcolor}
\usepackage{hyperref}
\usepackage{pdfcomment}
\usepackage{textalpha}

\newcommand{\sgn}{\operatorname{sgn}}
\newcommand{\rmd}{{\rm d}}

\begin{document}

\title{Geodesic completion of big bangs from emergent geometry}

\author{Brooke Berrios}
\author{Cameron Corley}
\author{Sky O'Donnell} 
\author{Benjamin Shlaer}
\email[Corresponding author: ]{bshlaer@calpoly.edu}
\author{Jada Young}
\affiliation{Department of Physics, California Polytechnic State University, San Luis Obispo, CA}

\begin{abstract}
Chaplygin gas and other k-essence models exhibit emergent geometry, with perturbations propagating on an acoustic metric disformally related to the Einstein-frame metric. For superluminal sound speed, we identify the disformal metric as the “causal frame,” since choosing a finite causal-frame lapse yields hyperbolic equations of motion for fields propagating in either frame. We show that with a phantom Chaplygin gas, the Einstein-frame lapse is forced to pass smoothly through zero and change sign while the causal-frame lapse remains positive. As a result, Einstein-frame degrees of freedom (including the scale factor) undergo spontaneous time-reversal while the Chaplygin gas evolves monotonically, enforcing a robust non-singular bounce even in the presence of additional matter canonically coupled to the Einstein frame.
\end{abstract}

\maketitle
We propose what may seem like an unobtrusive addition to general relativity, namely a perfect fluid whose energy density is bounded both above and below by an arbitrarily small parameter $\lambda$.  Surprisingly, this addition can drastically alter the causal structure of the universe:  Time itself dynamically reverses for all gravitational degrees of freedom, including the scale factor.  

The possibility of time-reversal was suggested by Moschella and Novello for a quartic k-essence model \cite{Moschella:2021pji}.  They
found that the crossing of branching surfaces in phase space could be interpreted either as a bounce or as time reversing.  Our Chaplygin gas model selects time-reversal as the only possibility.  Using the mostly-positive metric convention, it has a k-essence description of the phantom Dirac--Born--Infeld (DBI) type,
\begin{align}\label{eq:SpDBI}
S[\phi] =  \lambda\int\sqrt{-G}\sqrt{1-G^{\mu\nu}\partial_\mu\phi\partial_\nu\phi} \;\rmd^4x,
\end{align}
where $G_{\mu\nu}$ is called the Einstein-frame metric because it appears in the Einstein-Hilbert action. This action has a world-volume interpretation as a brane whose induced metric $g_{\mu\nu}$ is disformally related to Einstein frame via
\begin{align}\label{eq:causalmetric}
g_{\mu\nu} = G_{\mu\nu} - \partial_\mu\phi\partial_\nu\phi.
\end{align}
Notice that any vector that is null with respect to $G_{\mu\nu}$ is null or time-like with respect to $g_{\mu\nu}$, whose causal cone is thus {\em wider} \cite{Bekenstein:1992pj} than the light cone of the Einstein-frame metric $G_{\mu\nu}$. 
The deceptively simple-looking equation of motion for $\phi$ is harmonic in $g_{\mu\nu}$:
\begin{align}
\partial_\mu\left(\sqrt{-g}g^{\mu\nu}\partial_\nu\phi\right) = 0.
\end{align}
Scalar field perturbations propagate along the causal cones \cite{Babichev:2007dw} of $g_{\mu\nu}$, 
so phantom Chaplygin gas has a superluminal sound speed. The cure \cite{Bruneton:2006gf,Dubovsky:2005xd} for superluminality is to promote $g_{\mu\nu}$ to be the arbiter of causality, which we therefore call the {\em causal-frame metric}, and treat the Einstein-frame metric $G_{\mu\nu} =  g_{\mu\nu} + \partial_\mu\phi\partial_\nu\phi$ as an emergent geometry seen by gravitational waves, as well as any matter canonically coupled to $G_{\mu\nu}$.

The general Arnowitt--Deser--Misner (ADM) initial-value formalism with respect to $g_{\mu\nu}$ will appear in a companion publication \cite{Shlaer}.   In this Letter we work with the fluid description in a cosmological setting, where only the causal-frame lapse function $n = 1/\sqrt{-g^{tt}}$ is freely specified.  
Using capital letters to denote the Einstein-frame, the FLRW metric is
\begin{align}\label{eq:flrwE}
G_{\mu\nu}\rmd x^\mu\rmd x^\nu = -N^2(t)\rmd t^2 + a^2(t)\left[\rmd \chi^2 + S_k^2(\chi) \rmd\Omega_2^2\right].
\end{align}
Homogeneity and isotropy require $\phi = \phi(t)$, so the scale factor $a(t) = A(t)$ is frame-independent.  The two metrics differ only in their time-time components (lapse functions), with the causal-frame lapse function $n$ related to the Einstein-frame lapse function $N$ via
\begin{align}\label{eq:nsol}
    n(t) = \sqrt{N^2(t) + \dot\phi^2(t)},
\end{align}
with $\dot\phi \equiv \rmd\phi/\rmd t$. 
From this, $S[\phi] = \lambda\int n a^3\rmd t $, and using the odd-in-$N$ ADM volume form $Na^3\rmd t$ for Einstein frame, the gravitational energy density of the phantom DBI scalar is
\begin{align}\label{eq:rhodef}
    \rho_\lambda = -\lambda N/\sqrt{N^2 + \dot\phi^2} = -\lambda N/n,
\end{align} 
which is bounded above and below. The pressure is
\begin{align}\label{eq:eos}
p_\lambda = -\lambda^2/\rho_\lambda,    
\end{align}  
the defining equation of state for Chaplygin gas \cite{Kamenshchik:2001cp}. Since $w = p_\lambda/\rho_\lambda < -1$, it is called a {\em phantom} \cite{Caldwell:1999ew}, although the null-energy condition is not violated when $\rho_\lambda < 0$.

The Friedmann equations are 
\begin{align}
    H^2 &= \frac{8\pi G_{N}}{3}\rho_{\rm tot} - \frac{k}{a^2}\label{eq:friedmann1} ,\\
    \dot a &= N a H,\label{eq:adot1}\\
    \dot H &= -N\left(H^2 + \frac{4\pi G_N}{3} \left(\rho_{\rm tot} + 3 p_{\rm tot}\right)\right).\label{eq:Hdot}
\end{align}
The continuity equation $\nabla_\nu T^{\mu\nu} = 0$ for Chaplygin gas is 
\begin{align}\label{eq:rhodot}
    \dot \rho_\lambda = N\frac{3H\left(\lambda^2 -\rho_\lambda^2\right)}{\rho_\lambda},
\end{align}
which can be formally integrated to yield
\begin{align}\label{eq:rhoofa}
    \rho_\lambda = \pm\lambda\sqrt{1 - a_{\rm min}^6/a^6}.
\end{align}
Note $\rho_\lambda$ vanishes when the scale factor decreases to the integration constant $a_{\rm min}$.  Division by a vanishing energy density renders eq.~(\ref{eq:rhodot}) numerically singular.  However it is the causal frame lapse $n$ that is freely specifiable, and from eq.~(\ref{eq:rhodef}) the Einstein-frame lapse is determined via
\begin{align}\label{eq:Nsol}
    N(t) = -n(t)\rho_\lambda(t)/\lambda.
\end{align}
This solution for $N$ is important for two reasons.  First, the evolution eq.~(\ref{eq:rhodot}) for the Chaplygin gas becomes the completely regular expression 
\begin{align}\label{eq:rhodot2}
    \dot\rho_\lambda = -3nH\left(\lambda - \rho_\lambda^2/\lambda\right),
\end{align}
which is linear in $n$, validating Bruneton \cite{Bruneton:2006gf}.
Second, with an unambiguous, bounded expression for $N$, we can smoothly evolve every other equation of motion.  

We can diagnose the illness of vanishing $\rho_\lambda$ while attempting to freely specify $N > 0$: The causal frame lapse diverges, signaling a non-spacelike foliation of the causal frame.  The cure (specifying $n$) leads to a foliation that is spacelike for both frames \cite{Bruneton:2006gf,Babichev:2007dw}.

\begin{figure}[htbp]
\includegraphics[width=0.95\linewidth]{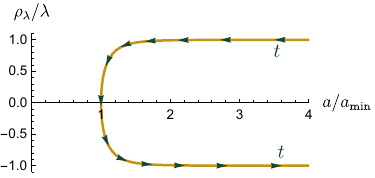} 
\caption{The Chaplygin gas energy density vs.~scale factor from eq.~(\ref{eq:rhoofa}).   The branch point where $\rho_\lambda$ changes sign represents a Big Brake singularity in Einstein frame, but is smoothly navigated in causal frame, as shown by the $t$-parameterized arrows.  For $H(t)>0$, causal frame cosmic time $t$ flows in the direction of decreasing $\rho_\lambda$, in agreement with eq.~(\ref{eq:rhosol}).   }
\label{fig:rho of a}
\end{figure}

Integrating eq.~(\ref{eq:rhodot2}) yields a sign-changing energy density
\begin{align}\label{eq:rhosol}
    \rho_\lambda(t) = -\lambda\tanh\left(3\int_{t_{\ast}}^tH\,n\rmd t'\right).
\end{align}
We choose ${n(t) = 1}$, and so the time coordinate $t$ is causal-frame cosmic time.
The behavior of eqs.~(\ref{eq:rhoofa}) and (\ref{eq:rhosol}) is shown in Fig.~\ref{fig:rho of a}.

While $\rho_\lambda(t)$ {\em smoothly} changes sign at $t = t_\ast$, the pressure diverges, signifying a sudden \cite{Barrow:2004xh} singularity  in Einstein frame similar to the Big Brake \cite{Gorini:2003wa}.  
The name Big Brake refers to the fact that the Einstein-frame deceleration parameter diverges at $t=t_\ast$ due to the pressure pole. 
However, in causal frame
the scale factor $a(t)$ smoothly bounces, but not because $H$ changes sign in eq.~(\ref{eq:adot1}), rather because the Einstein-frame lapse function ${N = -\rho_\lambda/\lambda}$ does.  
{\color{black}
Since $N(t)$ smoothly changes sign at $t = t_\ast$, the Einstein-frame metric momentarily degenerates at the crossing.  The causal frame remains non-singular, and $n=1$ provides a regular foliation that is everywhere spacelike for both frames.  All evolution equations (\ref{eq:adot1},\ref{eq:Hdot}, \& \ref{eq:rhodot2}) are regular through the $N=0$ crossing.}

The Einstein-frame Hubble parameter remains positive throughout, assuming $\lambda \ll \rho_{\rm tot}$.
This is important because it shows that all Einstein-frame-coupled fields smoothly experience time reversal, since their evolution equations are linear in $N$, e.g. in eqs.~(\ref{eq:adot1}-\ref{eq:Hdot}). The sole exception is $\rho_\lambda$ (equivalently $N$), whose velocity $\dot\rho_\lambda$ does not change sign when $N$ does, as shown in eq.~(\ref{eq:rhodot2}).

\begin{figure}[htbp]
\includegraphics[width=1.00\linewidth]{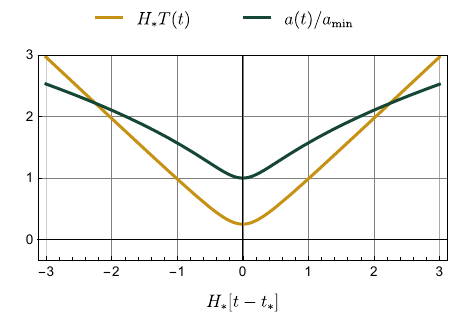}
\caption{Illustration of the time-reversal mechanism during radiation domination, with $H_\ast \equiv H(t_\ast) \gg \sqrt{G_N\lambda}$.  Einstein-frame cosmic time $T = \int N\rmd t$ attains a local minimum at causal-frame cosmic time $t = t_\ast$ when $N(t)$ changes sign.  For the same reason, the scale factor 
$a(t)$ exhibits a non-singular bounce without $H(t)$ changing sign.}
\label{fig:bounce}
\end{figure}

Unlike previous approaches that require distributional matter sources \cite{Keresztes:2012zn}, we have regularized the Big Brake singularity by choosing the {\em causal-frame} lapse function $n$ to be finite.  This implies the Einstein-frame lapse function smoothly changes sign, which means the $t$-derivative of the scale factor (and of every matter field canonically coupled to the Einstein-frame metric) similarly changes sign. 
As seen in Fig.~\ref{fig:bounce}, time-reversal is what prevents the scale factor from decreasing below $a(t_\ast) = a_{\rm min}$, rather than any gravitationally-repulsive matter source, or modified gravity.  
{\color{black}
Remarkably, the bounce caused by the sign change of $N(t)$ is a purely kinematic effect dependent on $\rho_\lambda$ crossing a surface in phase space, and not on its gravitational coupling $\lambda$, since we are not relying on $H(t)$ changing sign.  For this reason $\lambda$ can be arbitrarily small.} This Big Crunch avoidance mechanism is therefore robust against arbitrary additions of matter, e.g., isotropizing ultra-stiff ($w>1$) fluids \cite{Erickson:2003zm}.  

Non-singular bounces \cite{Brandenberger:2016vhg} are notoriously difficult to construct \cite{Creminelli:2016zwa}, but have been successfully engineered via modified gravity \cite{Arkani-Hamed:2003pdi, Easson:2011zy, Ijjas:2016tpn}, but see also \cite{Dobre:2017pnt}.

It might seem surprising that the sign of $N$ can have any physical content, since the metric is insensitive to it, and the sign of $\sqrt{-G}$ can be absorbed into a coordinate change.   The reason $\sgn(N)$ matters here is because
the sign of $\sqrt{-G}/\sqrt{-g}$ is coordinate invariant, and it reaches a branch point when $N/n \to 0$.  Since we fixed $n=1$, the monodromy at this branch point can be encoded in $\sgn(N)$ without affecting the metric.
This is accomplished by the oriented ADM measure $\sqrt{-G} \mapsto N\sqrt{\,\overline G\,} = Na^3$, which makes use of $\sgn(N)$ to distinguish forward evolution from reverse evolution.
Anytime $\sgn(N)$ makes an appearance, the coordinate-invariant relative time-orientation of the two volume-elements is being probed.

Unlike the Einstein--Hilbert term and Einstein-frame-matter Lagrangians, the phantom DBI Lagrangian does not encounter a branch point when $N\to0$ in eq.~(\ref{eq:nsol}). This explains why monodromy has no effect on the Chaplygin gas.

Because $N/n$ smoothly changes sign, all Einstein-frame-coupled degrees of freedom spontaneously reverse their velocities, while causal-frame degrees of freedom (just the Chaplygin gas) continue to evolve forward in time.  This implies a transient mismatch in {\em thermodynamic} arrows of time.  For this reason, the time-reversal event could reasonably be called a Big Benjamin Button.

{\bf Geodesic completeness of causal frame.}  As long as $\dot\phi \neq 0$, meaning $a_{\rm min} > 0$, there is no possibility of a big bang or big crunch singularity.  Instead, the scale factor smoothly bounces, leaving all curvature invariants of $g_{\mu\nu}$ finite \footnote{Our completeness statement refers to the Einstein-DBI-system (with minimally-coupled standard model).  Additional exotic matter could lead to Big Rip-type singularities.}.

According to the Borde--Guth--Vilenkin theorem, the absence of singularities does not guarantee geodesic completeness in cosmological models \cite{Borde:2001nh}.  Their argument is based on the fact that a null geodesic has an affine parameter $\sigma$ obeying
\begin{align}
   \int \rmd \sigma = \int a(t)n\rmd t.
\end{align}
Geodesic completeness requires the affine parameter to be extendable, meaning the right-hand-side must diverge as $t \to -\infty$.  
Past-eternal inflationary models are geodesically incomplete.  This has motivated asymptotically static pasts (emergent universe scenario \cite{Ellis:2002we}), in addition to bouncing and cyclic scenarios \cite{Brandenberger:2016vhg}.
Our lower bound $a(t) \geq a_{\rm min} > 0$ is enough to guarantee the divergence of the affine parameter as $t \to  -\infty$, and so past-completeness.
Thus the causal-frame is geodesically complete for conventional reasons:  because it is initially contracting before a non-singular bounce.

{\bf Geodesic completeness of Einstein frame.}  Since the initial value formalism takes place in causal frame, there is no need for the Einstein frame to be geodesically complete.  Of course, it too has scale factor bounded below by $a_{\rm min}>0$, and so no big bang or big crunch singularities.  But unlike the causal frame,
the Einstein-frame metric fails to have an inverse when $N = 0$, leading to a divergence of the Einstein-frame Ricci scalar.

This is because pressure contributes to the Ricci scalar, and pressure exhibits a simple pole divergence during time reversal.  The interpretation of pressure is via the work it does, i.e., the change in $\rho_\lambda$ per volume $a^3$.  It is because $\rho_\lambda$ continues to evolve while the scale factor momentarily freezes, as seen in Fig.~\ref{fig:rho of a}, that pressure must diverge.  One might expect geodesic incompleteness in Einstein frame. 

On the other hand, the Hamilton equations for canonical fields remain non-singular through time reversal, with all field velocities being proportional to $N$.
Furthermore, (in the absence of time-reversal) geodesics have been shown \cite{Fernandez-Jambrina:2004yjt, Barrow:2004xh} to be extendable through sudden singularities.  
We now show that point particle trajectories can in fact be unambiguously continued through moments of time reversal using the Hamiltonian flow.

{\bf Geodesics of the Einstein frame}.
We use the {\em oriented} point-particle action $S = -m\Delta\tau$, which is analogous to the ADM measure. This is the temporal {\em displacement} $\Delta \tau = \int \rmd \tau$ rather than the temporal distance $\int \sqrt{\rmd \tau^2}$.  
In terms of a path $x^\mu(\sigma)$ and the Einstein-frame metric and ADM measure, the temporal displacement is
\begin{align}\label{eq:deltatau}
\Delta \tau &= \int \sgn(N t_{,\sigma})\sqrt{-G_{\mu\nu}x^\mu_{,\sigma}x^\nu_{,\sigma}}\,\rmd\sigma,\\
&=\int N(t(\sigma))t_{,\sigma}\sqrt{1 - \left(\frac{a(t(\sigma))}{N(t(\sigma))}\frac{\chi_{,\sigma}}{t_{,\sigma}}\right)^2}\,\rmd\sigma,
\end{align}
where in the last line we have assumed constant $\theta(\sigma),\varphi(\sigma)$. 
Canonical momenta $P_t$ and $P_\chi$ obey the primary constraint
\begin{align}
    P_t + N\sqrt{m^2 + P_\chi^2/a^2} \approx 0,
\end{align}
which implies that $t_{,\sigma}$ is a freely-specifiable gauge choice.   We choose it to obey $t_{,\sigma}>0$.
The Hamiltonian is
\begin{align}
    H = t_{,\sigma}\left[P_t + N\sqrt{m^2 + P_\chi^2/a^2} \right].
\end{align}
The equation of motion for the comoving position $\chi(\sigma)$ is
\begin{align}
    \frac{a(t) \chi_{,\sigma}}{N(t) t_{,\sigma}} = \frac{P_\chi/a(t)}{\sqrt{m^2 + P_\chi^2/a^2(t)}},
\end{align}
where canonical momentum $P_\chi$ is the conserved comoving momentum.  Notice that the comoving velocity $\chi_{,\sigma}$ is proportional to $N(t(\sigma))$, and so smoothly reverses direction when  $N$ changes sign.
Thus Einstein-frame particle trajectories smoothly experience time reversal when using a smooth causal-frame cosmic time parameterization, e.g., $t(\sigma) = \sigma$.  

An {\em affine} parameterization $\sigma(\Sigma)$ leads to a tangent vector of constant magnitude, say $-m^2$, implying
\begin{align}\label{eq:affine}
    t_{,\Sigma} = \frac{\sqrt{m^2 + a^2\chi_{,\Sigma}^2}}{|N|} = \frac{\sqrt{m^2 + P_\chi^2/a^2}}{|N|},
\end{align}
and so the canonical momenta obey
\begin{align}
P_\chi = \sgn(N)a^2\chi_{,\Sigma}\qquad P_t(t) = -\sgn(N)N^2t_{,\Sigma}\;.
\end{align}
These can be assembled into a four-vector we dub the {\em anti-tangent vector}
\begin{align}
P^\mu = \sgn(N)x^\mu_{,\Sigma}\;.
\end{align}
If we define a geodesic to be a path that parallel transports its own anti-tangent vector, 
\begin{align}\label{eq:geo}
    x_{,\Sigma}^\mu\nabla_\mu P^\nu = 0,
\end{align}
we recover the usual geodesic properties of being affinely parameterized, and extremizing the oriented path length action.

The unexpected factor of $\sgn(N)$ compensates for the fact that the time-orientation of the path (according to the ADM measure) suddenly does not match the orientation of the affine parameter.  Unlike Einstein-frame cosmic time, the Einstein-frame affine parameter is not allowed to attain a local minimum, since a path must be a single-valued function.  
While Einstein-frame geodesics are continuous and unambiguous, they are not smooth: there is a cusp in ${t(\Sigma) \sim \Sigma/\sqrt{|\Sigma|}}$ and a kink in ${\chi(\Sigma)\sim|\Sigma|}$ at the moment of time reversal.  This follows the general pattern of Einstein-frame having (traversable) sudden singularities, while causal frame is entirely smooth.

From eq.~(\ref{eq:affine}), an Einstein-null geodesic has an affine parameter that obeys
\begin{align}
    \int\rmd\Sigma = \int a(t) \left|N(t)\right|\rmd t,
\end{align}
which diverges in the asymptotic past since ${a(t) \geq a_{\rm min}}$ and ${N(t) \equiv 0}$ is not a fixed-point of eqs.~(\ref{eq:eos}), (\ref{eq:Hdot}), (\ref{eq:Nsol}), \& (\ref{eq:rhodot2}).
Thus the Einstein-frame geometry is past geodesically complete even when the Einstein-frame Hubble parameter obeys $H_{\rm ave} > 0$ \cite{Borde:2001nh}.

\begin{acknowledgments}
We thank Evan Wooldridge, Matthew Imbriani, and Alex May for helpful conversations.  This work was supported by the Marrujo Foundation and the William and Linda Frost Foundation.
\end{acknowledgments}

\bibliography{references}

\end{document}